\documentclass[twocolumn,showpacs,preprintnumbers,amsmath,amssymb]{revtex4}

\usepackage{graphicx}
\usepackage{dcolumn}
\usepackage{bm}

\begin{document}

\title{Cascade of Quantum Phase Transitions in Tunnel-Coupled Edge 
States}

\author{I. Yang$^{1}$, W. Kang$^{1}$, 
K.W. Baldwin$^{2}$, L.N. Pfeiffer$^{2}$, and  K.W. West$^{2}$}

\affiliation{$^{1}$James Franck Institute and Department of Physics,
 University of Chicago, Chicago, Illinois 60637\\
 $^{2}$Bell Laboratories, Lucent Technologies, 600 
 Mountain Avenue, Murray Hill, NJ 07974}

\date{\today}

\begin{abstract}
    
We report on the cascade of quantum phase transitions exhibited by 
tunnel-coupled edge states across a quantum Hall line junction.
We identify a series of quantum critical points between 
successive strong and weak tunneling regimes in the zero-bias 
 conductance. Scaling analysis 
shows that the conductance near the critical magnetic fields $B_{c}$ 
is a function of a single scaling argument $|B-B_{c}|T^{-\kappa}$, 
where the exponent $\kappa = 0.42$. This puzzling resemblance to a
quantum Hall-insulator transition points to importance of interedge 
correlation between the coupled edge states.

\end{abstract}

\maketitle

Edge states in the quantum Hall effect provide a highly tunable system 
for the study of quantum transport in one-dimension\cite{KaneFisher}.  
Following the prediction of chiral Luttinger liquids in the fractional 
quantum Hall effect\cite{Wen90a,Wen91}, extensive effort has been 
devoted to the study of tunneling between quantum Hall edge 
states\cite{Kane92,Kane94,Fendley95,Chamon97,Milliken96,Chang96,
Grayson98,Shytov98}. 
Tunneling of an electron into a Luttinger liquid is strongly suppressed 
and theories predict a power-law tunneling conductance with a 
universal exponent related to the quantum number of the
bulk quantum Hall liquid.
Experimental studies of tunneling between edge states across 
a quantum  point contact\cite{Milliken96} and tunneling between an edge 
state and a three-dimensional metal\cite{Chang96,Grayson98} have 
generally tended to support the predicted Luttinger liquid behavior. 
However, there remain important open questions regarding the 
experimentally observed exponent and its correlation to the bulk 
quantum Hall states\cite{Shytov98}.

A different and perhaps more intriguing geometry for the study of 
edge state tunneling involves a line junction that juxtaposes two 
parallel, counterpropagating edge modes against each other.
Such a junction has been initially envisioned as a Hall bar with
a long narrow gate that couples two right and left moving edge channels 
of fractional quantum Hall liquids\cite{Renn95,Kane97}.
In the limit of weak bias, the conductance across the line junction 
remains quantized as backscattering between the edge states is negligible. 
For strong bias, inter-edge backscattering is suppressed and the 
conductance across the line junction vanishes. In between the two 
limits, the inter-mode backscattering is mediated by defects in the 
line junction and a metal-insulator transition is 
predicted\cite{Renn95,Kane97}. The transition is characterized by a 
temperature dependent conductivity that vanishes in the insulating 
phase and diverges in the metallic state in the limit of zero 
temperature.

Confirmation of the predicted metal-insulator transition has remained 
elusive as lithographic limitations and vertical offset of the gates 
from the plane of two-dimensional electrons complicate the realization 
of a line-junction. An alternate approach to a line junction involves 
taking advantage of the inherent atomic precision of molecular beam 
epitaxy (MBE) and inserting a precisely defined semiconductor barrier 
in the plane of two-dimensional electron system through the technique 
of cleaved edge overgrowth\cite{Pfeiffer90,Kang00}.
Such a junction strongly couples two counterflowing edge modes through 
interedge tunneling and features sharp resonances whenever the single 
particle energy levels coincide with the chemical potential\cite{Kang00}. 
These resonances are particularly enhanced in its width and height at 
zero-bias crossings, indicating the importance of electron-electron 
interaction.
Proposed explanations include enhanced tunneling driven by 
electron-electron interaction\cite{Mitra01,Lee01,Kollar02}, mixing of 
the states with equal transverse momentum from the opposite sides of 
the barrier\cite{Ho94,Takagaki00,Nonoyama02}, and a coupled Luttinger 
liquid interacting through a strongly backscattering center in the 
barrier\cite{Kim03}.

In this paper, we report on the observation of a cascade of quantum 
phase transitions exhibited by tunnel-coupled edge states of quantum 
Hall line junctions.
Two counterpropagating edge states are separated by an 8.8 nm-wide, 
$\sim$100$\mu$m-long semiconductor barrier. 
We identify a series of quantum critical points between 
successive strong and weak tunneling regimes that are reminiscent of 
the metal-insulator transition in two-dimensions. Scaling analysis 
shows that the conductance near the critical magnetic field $B_{c}$ 
is a function of a single scaling argument $|B-B_{c}|T^{-\kappa}$, 
where the exponent $\kappa \approx 0.42$. This apparent similarity to 
the quantum Hall-insulator transitions is quite puzzling due to
one-dimensional character of edge states. Whether the resemblance 
to a quantum Hall-insulator transition is coincidental or occurs 
from some deeper physics remains to be clarified.

\begin{figure}
\includegraphics[width=3.25in]{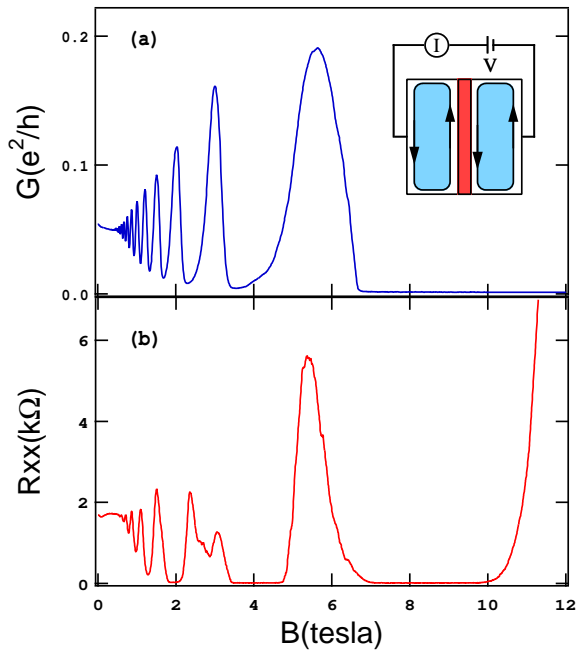}
\caption{\label{Fig:lnjnc}
(a) Representative zero-bias  conductance, $G = I/V$, of the line 
junction at 300 mK. Inset: Layout of the line junction and 
measurement geometry. Two counterpropagating edge states are 
juxtaposed against the barrier in the quantum Hall regime.
(b) Magnetoresistance from one of the two-dimensional 
electron system in the line junction. } 
\end{figure}

The line junctions are fabricated by cleaved edge overgrowth
using MBE\cite{Pfeiffer90,Kang00}. 
The initial growth on a standard (100) GaAs substrate consists of 
an undoped 13$\mu m$ GaAs 
layer followed by an  8.8 nm-thick digital alloy of undoped 
Al$_{0.1}$Ga$_{0.9}$As/AlAs, and completed by a 14$\mu m$ layer of 
undoped GaAs. This multilayer sample is cleaved along the (110) plane
in an MBE machine and a modulation-doping sequence is performed over 
the exposed edge, forming two strips of two-dimensional electron 
systems 
separated from each other by the 8.8 nm-thick barrier. A mesa 
incorporating the barrier and the two-dimensional electron systems 
into a junction that is $\sim$100$\mu m$ long is defined 
by photolithography. 
The inset of Fig. \ref{Fig:lnjnc}a shows the planar layout of the 
line junction device. In the quantum Hall regime, Landau quantization 
creates two counterpropagating edge states that are separated by a 
smooth, rectangular barrier. The density of the two-dimensional electron 
system in the devices studied 
was $n = 2\times 10^{11} cm^{-2}$ with a mobility of $\sim 1\times 
10^{5} cm^{2}/Vsec$.

Fig. \ref{Fig:lnjnc} illustrates the zero-bias  
conductance (ZBC), $G = I/V$, across the line junction and the 
magnetoresistance of the two-dimensional electron system
parallel to the tunnel barrier. The ZBC exhibits a 
series of conductance peaks that oscillates with increasing magnetic 
field before abruptly dropping to zero above 6.7 tesla. No oscillatory 
features can be seen at higher magnetic fields. This is thought to occur 
from decoupling of the counterpropagating edge modes beyond the last 
zero bias conductance peak\cite{Mitra01,Takagaki00,Nonoyama02,Kim03}. 
Shubnikov-de Haas oscillations are found in the magnetoresistance for 
low magnetic fields and integer quantum Hall states beyond 2 tesla.
The period of Shubnikov-de Haas oscillations of the two-dimensional 
electron systems does not match the conductance oscillations, which are  
sharper and more distinct than the oscillations in the 
magnetoresistance.  The mismatch in the 
oscillations arises naturally as a consequence of the electronic states near the 
junction occurring at higher energies than the corresponding bulk states 
with same quantum number\cite{Kang00,Ho94}. The single particle
energy levels near the barrier consists of a series of intersecting 
Landau levels from the left and right sides of the barrier.  The 
uncompensated carriers beneath the barrier further shifts the energy 
levels in vicinity of the tunnel barrier from that of an ideal 
two-dimensional electrons with uniform areal density\cite{Mitra01}.

In the noninteracting picture of tunneling across the line junction,
the ZBC peaks occur whenever the energy levels of the 
left and right edges coincide with the Fermi level at zero bias.
Under Landau quantization the spatial coordinate, $x$, 
corresponds to a guiding center state with a 
well-defined transverse momentum, $k_{y}$, through the relation 
$x = -k_{y}\ell_{\circ}^{2}$, where $\ell_{\circ}$ is the magnetic 
length.  In 
case of a high quality, low disorder barrier, tunneling across the 
junction must conserve momentum as the transverse 
momentum, $k_{y}$, is a good quantum number.
Whenever the levels coincide, states with equal transverse 
momentum are mixed and electrons from  one side of the barrier can 
tunnel over to the opposite side. The ZBC peaks consequently 
represent tunneling between edge states with $x = 0$ or conversely 
$k_{y} = 0$ momentum states. Since the  $x = 0$ guiding center state 
lies at the center of the barrier, there is a large overlap of 
the electronic wave functions which facilitates tunneling through the 
barrier. Introduction of electron-electron interaction creates a 
tunnel gap in the energy spectrum as the gain in the correlation energy 
compensates for the cost of Coulomb interaction energy\cite{Mitra01,Kollar02}. 
This is thought to be responsible for the enhancement in the width and 
height of the ZBC peaks.

\begin{figure}
\includegraphics[width=3.25in]{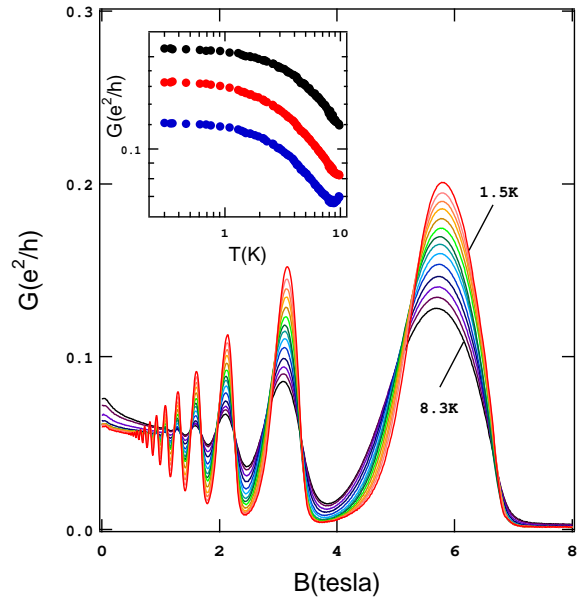}
\caption{\label{Fig:tdep}
Conductance of a line junction for various temperature 
between 1.3K and 8.5K.
Inset: Temperature dependence of  conductance the 
first 3 peaks. }
\end{figure}

Fig. \ref{Fig:tdep} shows the magnetic field dependence of the 
ZBC between 1.5 and 8.3K. 
The ZBC peaks grow in amplitude
with increasing magnetic field and decreasing temperature. 
Above 7 tesla, the ZBC becomes vanishingly small as the 
momentum conserved tunneling across the line junction can no longer 
be satisfied and the conduction occurs parallel to the junction, 
along the barrier. A striking feature of the conductance data 
in Fig. \ref{Fig:tdep} is the series of critical points on the high 
field side of the conductance peaks. These critical points separate 
the ZBC peaks from the low conductance regions where the tunneling 
is largely suppressed. Interestingly, no critical points can be seen 
on the lower field side of the ZBC peaks. In terms of single particle 
levels, there are excess states above the energy level crossings 
on the low field side prior to the entry into the zero-bias peaks. 
On the other hand, electronic states are depopulated as soon as 
the system exits the ZBC peaks on the high field side. Consequently,
the observed aymmetry may 
be reflecting the structure of the energy level crossings as the 
population of the filled states change as a function of magnetic 
field. 
The inset of Fig. \ref{Fig:tdep} illustrates the 
temperature dependence of ZBC at the three largest zero 
bias-conductance peaks. ZBC increases slowly as temperature is 
reduced and saturates below 1K. 

\begin{figure}
\includegraphics[width=3.25in]{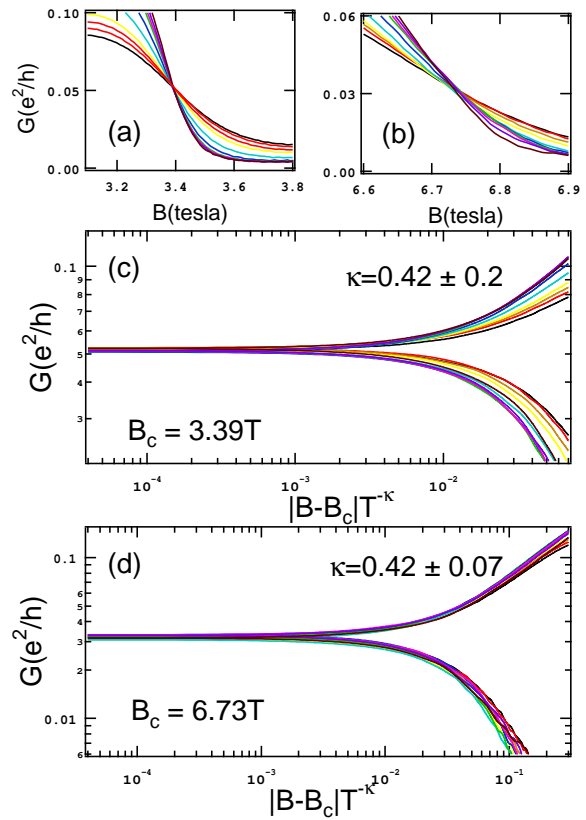}
\caption{\label{Fig:tran} 
Critical points and scaling analysis of the tunneling conductance. 
(a) Conductance data near $B_{c}$ = 3.39T. (b) Conductance data near 
$B_{c}$ = 6.73T. (c) Scaling analysis of the conductance data 
near $B_{c}$ = 3.39T as a function of $|B-B_{c}|T^{-\kappa}$. (d) 
Similar analysis performed for data near near $B_{c}$ = 6.73T.}
\end{figure}

Figs. \ref{Fig:tran}a and \ref{Fig:tran}b show an expanded view 
of the ZBC near the critical points around $B_{c}$ = 3.39T and 6.73T 
with corresponding critical conductance values of $G_{c}$ of 
approximately $0.05e^{2}/h$ and $0.03e^{2}/h$. 
Immediately above (below) the critical magnetic field, $B_{c}$,
ZBC decreases (increases) with temperature. Such a behavior about 
the critical points is reminiscent of the quantum Hall-insulator 
transitions in two-dimensions. Fig. \ref{Fig:tran}c and \ref{Fig:tran}d 
illustrate the results of the scaling analysis of the ZBC about 
the critical points.  For both 
cases we find that the tunneling conductance $G$ near $B_{c}$ can be 
scaled as an argument of $|B-B_{c}|T^{-\kappa}$, where $\kappa = 0.42$.
While the critical point at $B_{c}$ = 3.39T features a limited scaling 
regime and consequently a greater uncertainty in the value of the critical 
exponent, the extended scaling regime around $B_{c}$ = 6.73T and its
smaller variance of the exponent provide confidence on the scaling 
form. Remarkably, this is the same universal scaling form and the 
exponent found in quantum Hall-insulator transitions in bulk 
two-dimensional electron systems\cite{Sondhi97}.

The scaling result seen above raises a number of important questions 
regarding the quantum Hall line junctions: namely, (1) what is the 
physics behind the observed phase transitions, (2) what phases lie 
on either side of the critical points, and (3) what is the significance 
of the similarity to the quantum Hall-insulator transitions?  Answers 
to these questions are largely unknown and requires further 
theoretical investigation. Based on the separation of the edge states by 
a tunnel barrier on the order of magnetic length, it follows 
that correlation of electrons on the opposite sides of the barrier 
should play an important role in the electronic transport across the 
line junction.  The effect of electron-electron interaction is then to 
transform the pair of counterpropagating edge states into ground 
states characterized by the interedge Luttinger correlation. 

The high conductance and the low conductance regimes
then represent a pair of highly correlated ground states that are 
separated by a quantum phase transition and that differ primarily in 
its ability to tunnel across the line junction. 
The high conductance, ``metallic'' phase corresponds to a state 
with a number of edge electrons partaking in the tunneling across 
the barrier and  backscattering parallel to the junction. Above the 
critical magnetic fields, tunneling become suppressed and
primary conduction now occurs parallel to the barrier in
the low conductance, ``insulating'' phase. 
While the resemblance to a quantum Hall-insulator transition is
suggestive of some type of quantum Hall physics,  edge states are 
generally coupled weakly to the bulk quantum Hall states and are 
predominantly one-dimensional in its character. 
Whether these states possess significant enough quantum Hall 
correlation to produce the observed exponents remains unclear.

Although a disorder driven metal-insulator transition in a 
line junction has been predicted earlier\cite{Renn95,Kane97}, 
the high quality of the MBE-grown barrier and the momentum conservation 
in the single particle tunneling lead us to discount the likelihood 
of disorder playing a prominent role. The ballistic property of edge 
states further minimizes the possible decoherence effects associated
with disorder. 
Based on these features of the line junction, we conclude 
that disorder should not be playing an appreciable role in the 
observed transitions.

In the theory of tunneling based on the interedge phase 
coherence across the line junction, the ZBC peak states are 
explained in terms of a broken symmetry state characterized by 
a tunnel gap in the energy spectrum\cite{Mitra01}. Interaction 
between the left and the right edge states produces a Luttinger 
liquid whose symmetry is broken by a phase transition into a 
one-dimensional pseudospin ferromagnet. The gap in the tunnel 
spectrum estimated to be $\sim$1K in the samples with 8.8 nm-wide 
barrier\cite{Mitra01,Kollar02}. As the magnetic field is switched away from 
the ZBC peaks, the cost in the Coulomb energy increases as the 
tunnel gap is reduced. The system evolves continuously until it can no 
longer sustain a tunnel gap. Subsequent motion of electrons occurs
parallel to the barrier as tunneling is no longer possible.
It remains to be seen whether such a scenario will produce a 
quantum phase transition with observed critical exponents.

In the model of Kim and Fradkin\cite{Kim03}, inter-edge tunneling in 
the line junction
is equivalent to a coupled one-dimensional system interacting through 
short range interactions. Instead of considering a continuous distribution 
of tunneling sites along the junction, it is postulated that the tunneling 
between the right- an 
left-moving edge modes occurs primarily through a weak tunneling center.
Introduction of electron-electron interaction within the proposed 
framework allows for a rigorous mapping of two parallel edge channels 
into a coupled Luttinger liquid characterized by an 
effective Luttinger parameter $K$.
Depending on the coupling constant between the left and right moving 
branches, there is a quantum phase transition between a state with no
tunneling for $K > 1$ or perfect tunneling $K < 1$. The experimentally 
observed sequence of critical points represents a series of $K = 1$ 
quantum critical points between the strongly and weakly tunneling regimes. 
The sequence of critical point therefore mimics a series of opening 
and pinching-off of the tunneling center as a function of magnetic 
field.
While our data is qualitatively consistent with the proposed scenario 
by Kim and Fradkin, 
further clarification of the predicted transitions and associated 
critical exponents is necessary.

In conclusion, we have studied the temperature dependent transport 
across a quantum Hall line junction. The tunnel-coupled, 
counterpropagating edge states produce a series of quantum critical 
points between the highly and weakly tunneling regimes.
These critical points indicate a series of quantum 
phase transitions between two correlated one-dimensional ground 
states arising as a result of strong interedge correlation.
Scaling analysis shows that the conductance near the critical 
behavior scales as $|B-B_{c}|T^{-\kappa}, \kappa \approx 0.42$,  
similar to that of quantum Hall-insulator transitions. Whether there 
is strong quantum Hall correlation across the line junction or some 
other physics is responsible for the observed transitions remains to 
be explained.

We would like to thank E. Fradkin, S. Girvin, I. Gruzberg, E. Kim, A. Ludwig, 
A. Mitra, H.L. Stormer, and  P. Wiegmann for 
useful discussions. The work at the University of Chicago is supported by
NSF DMR-0203679 and NSF MRSEC Program under DMR-0213745.


\begin{thebibliography}{99}

\bibitem{KaneFisher} See review by C.L. Kane and M.P.A. Fisher in 
{\em Perspectives on Quantum Hall Effects}, 
 edited by S. Das Sarma and A. Pinczuk (Wiley, New York, 1997). 

\bibitem{Wen90a} X.G. Wen, Phys. Rev. B, {\bf 41}, 12838 (1990). 

\bibitem{Wen91} X.G. Wen, Phys. Rev. B, {\bf 43}, 11025 (1991). 

\bibitem{Kane92} C.L. Kane and M.P.A. Fisher, Phys. Rev. B {\bf 46},
15233 (1992); Phys. Rev. Lett. {\bf 71}, 4381 (1993). 

\bibitem{Kane94} C.L. Kane and M.P.A. Fisher, Phys. Rev. Lett. {\bf 72}, 724 
(1994). 

\bibitem{Fendley95} P. Fendley, A.W.W. Ludwig, and H. Saleur, Phys. Rev. B {\bf 
52}, 8934 (1995). 

\bibitem{Chamon97} C. de C. Chamon and E. Fradkin, Phys. Rev. B {\bf 56}, 
2012 (1997). 

\bibitem{Milliken96} F.P. Milliken, C.P. Umback, and R.A. Webb, Solid State
Comm. {\bf 97}, 309 (1996). 

\bibitem{Chang96} A.M. Chang, L.N. Pfeiffer, and K.W. West, Phys. Rev. 
Lett. {\bf 77}, 2538 (1996). 

\bibitem{Grayson98} M. Grayson, D.C. Tsui, L.N. Pfeiffer, K.W. West, and 
A.M. Chang, Phys. Rev. Lett. {\bf 80}, 1062 (1998). 

\bibitem{Shytov98} A.V. Shytov, L.S. Levitov, and B.I. Halperin, Phys. Rev. 
Lett. {\bf 80}, 141 (1998). 

\bibitem{Renn95} S.R. Renn and D.P. Arovas, Phys. Rev. B {\bf 51}, 
16832 (1995). 

\bibitem{Kane97} C.L. Kane and M.P.A. Fisher, Phys. Rev. B {\bf 56}, 15231 
(1997). 

\bibitem{Pfeiffer90} L.N. Pfeiffer, K.W. West, H.L. Stormer, 
J.P. Eisenstein, K.W. Baldwin, D. Gershoni, and J. Spector, 
Appl. Phys. Lett. {\bf 56}, 1697 (1990). 

\bibitem{Kang00} W. Kang, H.L. Stormer, K.B. Baldwin, L.N. Pfeiffer, and 
K.W. West, Nature {\bf 403}, 59-61 (2000). 

\bibitem{Mitra01} A. Mitra and S.M. Girvin, Phys. Rev. B, {\bf 64}, 41309 
(2001). 

\bibitem{Lee01} H.C. Lee and S.R.E. Yang, Phys. Rev. B {\bf 63}, 
193308 (2001). 

\bibitem{Kollar02} M. Kollar and S. Sachdev, Phys. Rev. B {\bf 65}, 121304 
(2002).

\bibitem{Ho94} Tin-Lun Ho, Phys. Rev. B {\bf 50}, 4524 (1994). 

\bibitem{Takagaki00} Y. Takagaki and K.H. Ploog, Phys. Rev. B {\bf 62}, 
3766 (2000). 
 
\bibitem{Nonoyama02} S. Nonoyama and G. Kirczenow, Phys. Rev. B {\bf 66}, 
155334 (2002). 

\bibitem{Kim03} E. Kim and E. Fradkin, Phys. Rev. B {\bf 67}, 45317 (2003).

\bibitem{Sondhi97} S.L. Sondhi, S.M. Girvin, J.P. Carini, and D Shahar, 
Rev. Mod. Phys. {\bf 69}, 315 (1997). 


\end{thebibliography}
\end{document}